\documentclass[twoside,11pt]{article}

\usepackage[preprint]{jmlr2e}
\usepackage{blindtext}
\usepackage{amsmath}
\usepackage{multirow}
\usepackage{graphicx}
\usepackage{booktabs}
\usepackage{caption}
\usepackage{subcaption} 
\usepackage{algorithm}
\usepackage{algpseudocode}

\usepackage{lastpage}
\jmlrheading{}{2025}{}{}{}{}{Daniel Krasnov and David Stephens}

\ShortHeadings{Bayesian Nonparametric Detection of Anomalies in Multivariate Functional Data}{Krasnov and Stephens}
\firstpageno{1}

\begin{document}

\title{Bayesian Nonparametric Detection of Anomalies in Multivariate Functional Data}

\author{\name Daniel Krasnov \email daniel.krasnov@mail.mcgill.ca \\
       \addr Department of Mathematics and Statistics\\
       McGill University\\
       Montreal, BC H3A 0B9, Canada
       \AND
       \name David Stephens \email david.stephens@mcgill.ca \\
        \addr Department of Mathematics and Statistics\\
       McGill University\\
       Montreal, BC H3A 0B9, Canada}

\editor{}

\maketitle

\begin{abstract}
Anomalies in functional data arise from rare or distinct processes that deviate from the dominant data-generating mechanism. Detecting such departures is essential in applications where they may correspond to errors, structural changes, or other behavior of interest. This work introduces a Bayesian nonparametric approach for anomaly detection in multivariate functional data. We model functional data as an infinite mixture of multi-output Gaussian processes, with a finite and automatically determined number of mixture components obtained through slice sampling. Mean functions are represented using a wavelet basis and regularized through Besov priors to obtain a smooth and sparse representation of the data. Cross-functional dependence is captured using the intrinsic coregionalization model and we solve covariance kernel selection by introducing a Carlin-Chib product space step in the Markov Chain Monte Carlo algorithm. Within this model, anomalous observations are assigned to small mixture components without requiring prior specification of the number or nature of anomalies. We consider a semi-supervised setting, in which labels are available for 15\% of the normal observations and a large class imbalance is present. The utility of our model is demonstrated on both univariate and multivariate functional data.
\end{abstract}


\section{Introduction}

Functional data analysis (FDA) is the branch of statistics concerned with analyzing data that are most naturally viewed as functions. Typically, this means infinite-dimensional smooth curves defined over a continuous domain such as time or space \citep{ramsay2005functional}. FDA has proven to be effective across a wide array of fields such as biostatistics, environmetrics, linguistics, finance, psychology, meteorology, and engineering \citep{ullah2013applications}. In any of these application areas, we may encounter situations where we are interested in identifying unusual or uncommon signals. 

Functional anomaly detection (FAD) and functional outlier detection (FOD) involve the study of functional observations that deviate from the dominant behavior of the data. Detecting anomalous functions is important because, as noted by \citet{hawkins1980identification}, outliers can bias parameter estimates, obscure genuine relationships, and result in misleading conclusions. In general, FAD problems share two defining characteristics. First, there is typically a large class imbalance in the data: anomalous observations are rare and represent unusual and unwanted behavior. Second, the goal is to identify the underlying structure in the data, so as to classify observations as either normal or anomalous. This means we require some notion of normalcy and must encode it in the model. This is complicated by the fact that FAD problems are usually unsupervised, meaning that we have no prior knowledge of what anomalies look like nor how many distinct types may occur. Beyond detection, it is often desirable to understand the impact of these anomalies on the system under study. As a result, we also seek to estimate accurately quantities such as the mean, slope, phase, or variance. Finally, our specific setting of FDA adds one additional challenge as our data are assumed to be infinite-dimensional objects that vary with respect to some domain. This means that there could be complex forms of dependence both within and across the function's dimensions.

As outlined by \citet{staerman_functional_2023}, FAD methods can be grouped into three main categories: depth-based, projection-based, and model-based approaches. 
Depth-based methods quantify the centrality of each function within the sample through a depth function, ranking observations from the most central to the most outlying. 
Projection-based methods reduce functional data to a finite-dimensional representation using a chosen set of basis functions, after which machine learning techniques may be applied to detect deviations from the dominant structure. 
Model-based methods assume an explicit probabilistic model for the data-generating process and classify functions as anomalous based on the model's output. 

This work chooses to follow a model-based approach for FAD. Specifically, we adopt a Bayesian nonparametric (BNP) approach that models the joint distribution of functional observations using an infinite mixture of intrinsic coregionalization models (ICMs; \citealp{alvarez2011computationally}). As is shown in later Sections, anomalies arise as small mixture components that are combined at the end of our Markov Chain Monte Carlo (MCMC) algorithm to form an anomaly class. This approach eliminates the need for user-defined thresholds, provides uncertainty quantification through the Bayesian lens, and captures multivariate dependencies and nonstationary structures. Our main contributions are summarized as follows:

\begin{itemize}

\item[(i)] We introduce a Bayesian nonparametric model for anomaly detection in multivariate functional data, termed the Wavelet Intrinsic Coregionalization Model Anomaly Detector (WICMAD). The method models functional observations as an infinite mixture of multi-output Gaussian processes, allowing anomalous curves to emerge as small mixture components without requiring user-specified thresholds or prior knowledge of the number or structure of anomalies.
\item[(ii)] To capture complex functional structure, we combine wavelet-based mean modeling with a multi-output Gaussian process residual model based on the intrinsic coregionalization model. This decomposition enables flexible modeling of both local functional features and cross-dimensional dependence. Our method can be seen as a multivariate extension of \cite{ray_functional_2006}.
\item[(iii)] We develop a Markov chain Monte Carlo algorithm that incorporates a Carlin--Chib product-space move for automatic, cluster-specific covariance kernel selection. This allows the model to adaptively choose among multiple kernel families, reducing sensitivity to kernel misspecification.

\end{itemize}

The remainder of this article is organized as follows. 
Section~\ref{sec:meth} introduces the notation and theoretical framework underlying the proposed method, Section~\ref{sec:comp} describes the computational details of the MCMC algorithm, Section~\ref{chap:simstudy} presents the simulation study, Section~\ref{chap:apps} showcases the model's performance on real data, and Section~\ref{sec:conclu} concludes.

\section{Methodology}
\label{sec:meth}

This section introduces the proposed model. 
Section~\ref{subsec:prelim} formally presents the problem setting, 
Section~\ref{subsec:wavelet} describes the wavelet model for functional means, 
Section~\ref{sec:icm} introduces the intrinsic coregionalization model used to capture dependence across dimensions of the residual process, and
Section~\ref{sec:gp-kernels} discusses automatic kernel selection via a Carlin--Chib product-space move within the MCMC algorithm.

\subsection{Preliminaries}
\label{subsec:prelim}

Following \citet{staerman_functional_2023}, FDA considers data that are realizations of random functions
\[
X : (\Omega, \mathcal{A}, \mathbb{P}) \to \mathcal{F},
\]
where \( (\Omega, \mathcal{A}, \mathbb{P}) \) is the probability space and $X$ takes on values in a topological function space \( \mathcal{F} \). For the purposes of this article, \(\mathcal{F}\) will always be taken to be a Hilbert space, specifically the space of square-integrable functions \(L^2(\mathcal{T})\),
where \(\mathcal{T} \subset \mathbb{R}\) denotes the domain. We assume the data we observe exhibit noise according to

\[
Y_{ij} = X_i(t_{ij}) + \varepsilon_{ij},
\]
where \( X_i(\cdot) \in L^2(\mathcal{T}) \) denotes the \(i\)th function observed over the domain \(\mathcal{T}\), and \( \varepsilon_{ij} \) represents measurement noise. In the multivariate case,
\[
\mathbf{X}(t) = \big(X^{(1)}(t), \ldots, X^{(D)}(t)\big)^{\top}, \quad
\mathbb{E}\!\left[\int_{\mathcal{T}} \|\mathbf{X}(t)\|^2 dt \right] < \infty,
\]
where dependencies within and across dimensions are described by the mean function \( \boldsymbol{\mu}(t) = \mathbb{E}[\mathbf{X}(t)] \) and covariance function \( \boldsymbol{\Sigma}(s,t) = \operatorname{Cov}(\mathbf{X}(s), \mathbf{X}(t)) \).

FAD seeks to identify functional observations that deviate from the dominant structure of the data. This amounts to estimating the mapping
\[
g : \mathcal{X} \to \{-1, +1\},
\]
where \(\mathcal{X}\) is the set of our data, \( g(x) = 1 \) denotes an anomalous function and \( g(x) = -1 \) denotes a normal one. Throughout this work we will assume a semi-supervised setting wherein 15\% of the normal group's labels are available to the model and 15\% of the datasets are anomalies.

\subsection{Wavelet Mean Modeling}
\label{subsec:wavelet}

To model cluster-level structure flexibly, we represent the mean functions in a wavelet basis. After applying the discrete wavelet transform to the data \citep{mallat2002theory}, the mean function for dimension $m$ in cluster $k$ can be written as
\[
\mu_{k,m}(t)
= \sum_{\ell=1}^{L} \alpha_{k,m,\ell}\,\phi_{\ell}(t)
+ \sum_{j=1}^{J}\sum_{\ell=1}^{L_j}\beta_{k,m,j,\ell}\,\psi_{j,\ell}(t),
\]
where $\phi_\ell(t)$ are scaling functions and $\psi_{j,\ell}(t)$ are wavelet functions at scale $j$ and location $\ell$.

Following \citet{ray_functional_2006}, sparsity is induced through a Besov prior on the detail coefficients. Specifically,
\[
\beta_{k,m,j,\ell} \mid \gamma_{k,m,j,\ell}
\sim (1-\gamma_{k,m,j,\ell})\,\delta_0
    + \gamma_{k,m,j,\ell}\,\mathcal{N}\!\big(0,\,\sigma_{k,m}^2 g_{k,j}\big),
\qquad
\gamma_{k,m,j,\ell} \sim \mathrm{Bernoulli}(\pi_{k,j}).
\]
Here $\gamma_{k,m,j,\ell}$ indicates whether the coefficient is active, $\sigma_{k,m}^2$ governs dimension-specific variability, and $g_{k,j}$ controls the scale-dependent variance. The inclusion probabilities $\pi_{k,j}$ control the sparsity at each wavelet level and are assigned Beta priors,
\[
\pi_{k,j} \sim \mathrm{Beta}\!\big(\tau_\pi m_j,\;\tau_\pi (1-m_j)\big),
\qquad
m_j = \kappa_\pi\, 2^{-c_2 j}.
\]
Under this parameterization, $\mathbb{E}[\pi_{k,j}] = m_j$, so that finer scales have smaller expected inclusion probabilities. The hyperparameter $\tau_\pi > 0$ controls concentration around the mean, $\kappa_\pi \in (0,1)$ sets the expected inclusion probability at the coarsest scale, and $c_2 > 0$ determines the rate at which sparsity increases with resolution. As $m_j$ decays with $j$, coefficients at finer scales are a priori more likely to be inactive.

The parameter $c_2$ corresponds to one of the Besov smoothness parameters discussed by \citet{ray_functional_2006}. In their formulation, inclusion probabilities are fixed, making the choice of $c_2$ particularly important. Thus, they propose an estimation procedure to guide its selection. However, in our model we allow the $\pi_{k,j}$ to be random, which enables sparsity to be adaptively learned from the data, reducing sensitivity to this choice. Nonetheless, since $c_2$ still influences the implied Besov space, its selection may warrant additional consideration in certain applications. Further discussion can be found in \citet{ray_functional_2006}.

Finally, we discuss how wavelet variance is modeled. Each cluster $k$ includes a global scaling parameter $\tau_{\sigma,k}$ that controls dimension-specific variances. Conditional on this parameter, the variance for each dimension $m$ is modeled as
\[
\sigma_{k,m}^2 \mid \tau_{\sigma,k} \sim \mathrm{Inv\text{-}Gamma}\!\big(a_\sigma,\; b_\sigma\,\tau_{\sigma,k}\big),
\quad\text{with}\quad
\tau_{\sigma,k} \sim \mathrm{Gamma}(a_\tau, b_\tau).
\]

\subsection{Intrinsic Coregionalization Model}
\label{sec:icm}

This section discusses our use of the intrinsic coregionalization model (ICM) which represents the residual process within each cluster as a multi-output Gaussian process. For a full introduction to this model, see \citet{alvarez2011computationally}. Let $Y_{k,\mathrm{res}}\in\mathbb{R}^{P\times D}$ denote the residuals for cluster $k$, where $P$ is the number of time points and $D$ the number of dimensions. Vectorizing column-wise, the model assumes
\[
\operatorname{vec}(Y_{k,\mathrm{res}}) \sim \mathcal{N}\!\left(0,\;
K_x^{(s_k)} \otimes (\tau_{B,k} B_k^{\text{shape}}) \,+\, I_P \otimes \mathrm{diag}(\eta_k)\right),
\]
where $K_x^{(s_k)}$ is the $P\times P$ covariance matrix corresponding to the selected kernel family $s_k$, $B_k^{\text{shape}}$ is a $D\times D$ positive semi-definite coregionalization matrix with fixed trace, $I_P$ is the $P$-dimensional identity matrix, and $\eta_k\in\mathbb{R}_{+}^D$ contains dimension-specific noise variances. 

The above model is a slight deviation from the standard ICM.  In order to separate the overall scale of cross-dimensional dependence from its correlation structure, we have introduced an amplitude parameter $\tau_{B,k}>0$ such that the coregionalization matrix becomes
\[
B_k = \tau_{B,k}\,B_k^{\text{shape}},
\qquad \mathrm{tr}(B_k^{\text{shape}})=D.
\]
We further parameterize the shape matrix via a Cholesky decomposition,
\[
B_k^{\text{shape}} = D\,\frac{LL^\top}{\mathrm{tr}(LL^\top)},
\]
which enforces positive definiteness and resolves scale non-identifiability. Similar constraint strategies are standard in the multi-output GP literature; see, for example, \citet{bonilla2007multi,alvarez2011computationally,touloumis2021hypothesis}.

\subsection{Kernel Selection and Updating}
\label{sec:gp-kernels}

Employing the ICM requires us to select a covariance kernel for the model. The chosen kernel has large implications for model performance, as misspecification can result in anomalies being grouped with normal observations or the normal class being split into arbitrarily many clusters. It is therefore important to include a sufficiently rich and diverse set of kernels capable of capturing the range of behaviors that may arise in functional data. In particular, different anomaly types may exhibit varying degrees of smoothness, from non-differentiable to infinitely differentiable behavior, as well as structural features such as changepoints or nonstationary covariance patterns. To address this, we employ a Carlin--Chib product space move within the MCMC algorithm for automatic kernel selection \citep{carlin1995bayesian,lodewyckx2011tutorial}.

Let $s_k\in\{1,\dots,R\}$ denote the kernel--family indicator for cluster $k$, which corresponds to one covariance family from
\[
\mathcal K =
\Big\{
k^{(\mathrm{SE})},\;
k^{(\mathrm{Mat}\,3/2)},\;
k^{(\mathrm{Mat}\,5/2)},\;
k^{(\mathrm{Per})},\;
k^{(\mathrm{RQ})},\;
k^{(\mathrm{PowExp})},\;
k^{(\mathrm{GibbsPC3})},\;
k^{(\mathrm{CP\text{-}M52})}
\Big\}
\]
which respectively denote the squared exponential, Matérn with $\nu=3/2$, Matérn with $\nu=5/2$, periodic, rational quadratic \citep{rasmussen2006gaussian}, powered exponential \citep{diggle1998model}, Gibbs with piecewise--constant lengthscale in three bins \citep{paciorek2003nonstationary}, and changepoint Matérn--$5/2$ kernels \citep{lloyd2014automatic}, respectively. Each family $r$ has its own hyperparameters $\phi_{k,r}$, and all kernels are normalized to have unit marginal variance for identifiability.

Kernel switching proceeds as follows. For each cluster $k$ and each kernel family $r$, we define a parameter vector $\phi_{k,r}$ and refresh it from its pseudoprior when that kernel is inactive. The pseudopriors are chosen to match the corresponding priors, which simplifies the MCMC updates \citep{lodewyckx2011tutorial}. Only the active pair $(s_k,\phi_{k,s_k})$ contributes to the likelihood, and the indicator $s_k$ is sampled from a categorical distribution with weights
\[
w_r \propto
\mathcal{N}\!\Big(
  \operatorname{vec}(Y_{k,\mathrm{res}})\mid 0,\,
  K_x^{(r)}(\phi_{k,r}) \otimes \big(\tau_{B,k}\,B_k^{\text{shape}}\big)
  + I_P \otimes \mathrm{diag}(\eta_k)
\Big),
\]
under a uniform prior over $s_k$.

After selecting a covariance family, its hyperparameters are updated using random--walk Metropolis--Hastings \citep{sherlock2010random}. Positive parameters are proposed on the log scale, and parameters constrained to an interval are updated through an appropriate reparameterization.

\section{Implementation}
\label{sec:comp}

Posterior inference for the proposed model is performed using a Markov chain Monte Carlo (MCMC) algorithm that combines Gibbs sampling and Metropolis--Hastings updates. The sampler alternates between four main blocks of parameters:

\begin{enumerate}
\item cluster assignments and Dirichlet process weights,
\item wavelet mean parameters,
\item covariance kernel indicators and kernel hyperparameters,
\item ICM covariance parameters.
\end{enumerate}

Whenever conjugate priors are available, parameters are updated using Gibbs sampling. Non--conjugate parameters, including Gaussian process hyperparameters and ICM covariance parameters, are updated using random--walk Metropolis--Hastings proposals. Derivations of the conditional posterior distributions used in these updates are provided in Appendix~\ref{app:post}.

Section~\ref{sec:waveletmeanupdate} discusses Gibbs updates for the wavelet mean model, Section~\ref{sec:kernelupdate} discusses exploration of the space of kernels and their hyperparameter updates, and Section~\ref{sec:icmupdates} discusses ICM parameter updates. The main computational steps are summarized in Algorithms~\ref{alg:wavelet-gibbs}--\ref{alg:icm-mh}.

\subsection{Wavelet Mean Updates}
\label{sec:waveletmeanupdate}

Wavelet coefficients are updated independently within each cluster using conjugate Gibbs steps, with posterior details provided in Appendix~\ref{sec:wavelet_post}. Let $\mathcal{I}_k$ denote the set of curves assigned to cluster $k$. For each dimension $m$, wavelet scale $j$, and location $\ell$, we observe coefficients $y_{i,m,j,\ell}, i \in \mathcal{I}_k$. Sparsity is induced through a Besov prior with binary inclusion indicators $\gamma_{k,m,j,\ell}$. Each indicator is sampled from its Bernoulli posterior, determining whether the corresponding coefficient is active. If $\gamma_{k,m,j,\ell}=1$, the shared cluster coefficient $\beta_{k,m,j,\ell}$ is drawn from its Gaussian posterior; otherwise, it is set to zero. Scaling coefficients are always active and are sampled directly from their conjugate Normal posterior.

After updating the coefficients, we update the associated hyperparameters via conjugate Gibbs steps. The inclusion probabilities $\pi_{k,j}$ are sampled from Beta posteriors, the slab variance parameters $g_{k,j}$ from inverse--Gamma distributions, and the dimension-specific noise variances $\sigma_{k,m}^2$ from inverse--Gamma posteriors based on wavelet-domain residuals. Finally, the global variance scaling parameter $\tau_{\sigma,k}$ is sampled from its Gamma posterior. Algorithm~\ref{alg:wavelet-gibbs} summarizes the full Gibbs update for the wavelet model within a single cluster.

\begin{algorithm}[H]
\caption{Wavelet–block Gibbs updates for cluster $k$}
\label{alg:wavelet-gibbs}
\begin{algorithmic}[1]
\State Compute forward DWTs for all curves and precompute $\bar y_{m,j,\ell}$. Let $\mathcal{I}_k$ be the set of all indices in cluster $k$.
\For{$m=1,\dots,D$}
  \For{each \text{detail} level $j$ and location $\ell$}
    \State Compute $S_{1,m,j,\ell}=\sum_{i\in\mathcal{I}_k} y_{i,m,j,\ell}$ \text{at level $j$ when updating $\gamma$}.
    \State Sample $\gamma_{k,m,j,\ell}$ using the posterior inclusion probability built from $S_{1,m,j,\ell}$.
    \If{$\gamma_{k,m,j,\ell}=1$}
      \State Sample $\beta_{k,m,j,\ell}\sim\mathcal N(\mu^\star,v^\star)$
    \Else
      \State Set $\beta_{k,m,j,\ell}=0$
    \EndIf
  \EndFor
\EndFor
\State For each \text{scaling} coefficient $(m,s_J)$: set $\gamma_{k,m,s_J}=1$ and sample $\beta_{k,m,s_J}\sim \mathcal N\!\big(\bar y_{m,s_J},\,\sigma^2_{k,m}/|\mathcal{I}_k|\big)$.
\For{each detail level $j$}
  \State Sample $g_{k,j}$ and $\pi_{k,j}$ from their conjugate posteriors.
\EndFor
\For{$m=1,\dots,D$}
  \State Form wavelet-domain residuals $r_{i,m,j,\ell}=y_{i,m,j,\ell}-\beta_{k,m,j,\ell}$ and sample $\sigma_{k,m}^2$ from its inverse–Gamma posterior.
\EndFor
\State Sample $\tau_{\sigma,k}$ from its Gamma posterior.
\end{algorithmic}
\end{algorithm}

\subsection{Kernel Selection and Hyperparameter Updates}
\label{sec:kernelupdate}

Kernel selection is performed using a Carlin--Chib product-space formulation; full posterior derivations are provided in Appendix~\ref{sec:kernel_updates}. For each cluster $k$, let $s_k \in \{1,\dots,R\}$
denote the active covariance kernel family. At each iteration, parameters associated with inactive kernels are refreshed from their pseudopriors. The kernel indicator $s_k$ is then sampled from a categorical distribution with weights proportional to the marginal likelihood of the residual data under each candidate kernel,
\[
w_r \propto
\mathcal N
\Big(
\mathrm{vec}(Y_{k,\mathrm{res}})\mid 0,\,
K_x^{(r)}(\phi_{k,r}) \otimes (\tau_{B,k} B_k^{\text{shape}})
+ I_P \otimes \mathrm{diag}(\eta_k)
\Big).
\]
Conditional on the selected kernel, its hyperparameters are updated using random--walk Metropolis--Hastings steps.

In our setting, the kernel index $s_k$ is treated as a nuisance parameter rather than an object of direct inference. Our primary goal is to identify a kernel that provides a good fit to the data, rather than to characterize the full posterior distribution over the kernel space. To this end, we run multiple MCMC chains and select the model corresponding to the highest observed likelihood. We monitor the number of kernel switches within each chain and ensure that burn-in is chosen such that kernel exploration has stabilized before collecting posterior samples. Algorithm~\ref{alg:kernelswitch} summarizes the kernel update procedure.

\begin{algorithm}[H]
\caption{Kernel-family switch (Carlin--Chib) and MH update for cluster $k$}
\label{alg:kernelswitch}
\begin{algorithmic}[1]
\State \textbf{Inputs:} residuals $Y_{k,\mathrm{res}}\in\mathbb{R}^{P\times D}$; current $(s_k,\{\phi_{k,r}\}_{r=1}^R)$; $B_k^{\text{shape}}$, $\tau_{B,k}$, $\eta_k$; priors $p_r(\cdot)$
\State For each $r\neq s_k$: sample inactive parameters $\phi_{k,r}\sim q_r(\cdot)$ with $q_r=p_r(\cdot)$.
\For{$r=1,\dots,R$}
  \State Form $K_x^{(r)} \gets K_x^{(r)}(\phi_{k,r})$.
  \State $w_r \propto \mathcal{N}\!\big(\mathrm{vec}(Y_{k,\mathrm{res}})\mid 0,\ \tau_{B,k} B_k^{\text{shape}}\!\otimes\! K_x^{(r)} + \mathrm{diag}(\eta_k)\!\otimes\! I_P\big)\cdot \pi(s_k{=}r)$.
\EndFor
\State Draw $s_k \sim \mathrm{Categorical}\big(\{w_r\}_{r=1}^R\big)$.
\State Set $\phi \gets \phi_{k,s_k}$ and evaluate $\mathcal{L}(\phi)$ under $K_x^{(s_k)}(\phi)$.
\State Propose $\phi' \sim q(\cdot\mid \phi)$ using the family-specific rules.
\State Compute $\alpha \gets \min\!\left\{1,\ \frac{\mathcal{L}(\phi')\,p_{s_k}(\phi')\,q(\phi\mid\phi')}{\mathcal{L}(\phi)\,p_{s_k}(\phi)\,q(\phi'\mid\phi)}\right\}$; accept with probability $\alpha$.
\end{algorithmic}
\end{algorithm}

\subsection{Intrinsic Coregionalization Model Updates}
\label{sec:icmupdates}

To evaluate the likelihood efficiently, we exploit the separable structure of the covariance matrix. Let
\[
K_x^{(s_k)} = V \Lambda V^\top
\]
denote the eigendecomposition of the kernel matrix associated with the active kernel $s_k$. Projecting the residuals into this eigenbasis yields
\[
\tilde Y = V^\top Y_{k,\mathrm{res}},
\]
which decouples the likelihood across the $P$ transformed components. Under this transformation, the covariance of each component takes the form
\[
C_i = \tau_{B,k}\lambda_i B_k^{\text{shape}} + \mathrm{diag}(\eta_k), \quad i = 1,\dots,P,
\]
where $\lambda_i$ denotes the $i$th eigenvalue of $K_x^{(s_k)}$. The resulting log-likelihood can then be written as
\[
\ell(\tau_{B,k}, B_k^{\text{shape}}, \eta_k)
=
-\frac12
\sum_{i=1}^{P}
\left(
\log |C_i|
+
\tilde y_i^\top C_i^{-1} \tilde y_i
\right),
\]
where $\tilde y_i$ denotes the $i$th column of $\tilde Y$. The parameters $(\tau_{B,k}, B_k^{\text{shape}}, \eta_k)$ are updated using random--walk Metropolis--Hastings proposals, with full details provided in Appendix~\ref{app:icmupdates}. Algorithm~\ref{alg:icm-mh} summarizes the complete update procedure.

\begin{algorithm}
\caption{ICM Metropolis--Hastings updates for cluster $k$}
\label{alg:icm-mh}
\begin{algorithmic}[1]
\State \textbf{Inputs:} $Y_{k,\mathrm{res}}\in\mathbb{R}^{P\times D}$, active kernel $K_x^{(s_k)}(\phi_{k,s_k})$, current $(\tau_{B,k},B_k^{\text{shape}},\eta_k)$
\State Eigendecompose $K_x^{(s_k)}=V\Lambda V^\top$, set $\tilde Y\gets V^\top Y_{k,\mathrm{res}}$.
\State For $i=1,\dots,P$, set $C_i \gets \tau_{B,k}\lambda_i B_k^{\text{shape}}+\mathrm{diag}(\eta_k)$ and let $\tilde y_i\in\mathbb{R}^D$ be row $i$ of $\tilde Y$.
\State Define $\ell(\tau_{B,k},B_k^{\text{shape}},\eta_k) \gets -\tfrac{1}{2}\sum_{i=1}^P\!\big(\log|C_i|+\tilde y_i^\top C_i^{-1}\tilde y_i\big)$.
\Statex
\State \text{Amplitude update} ($\tau_{B,k}$): propose $\log\tau_{B,k}'=\log\tau_{B,k}+\epsilon_\tau$, $\epsilon_\tau\sim\mathcal{N}(0,s_\tau^2)$
\State \hspace{1.2em}Acceptance:
\[
\alpha_\tau=\min\!\left\{1,\;
\exp(\ell'-\ell)\,
\frac{p(\tau_{B,k}')}{p(\tau_{B,k})}\,
\frac{q(\tau_{B,k}\mid\tau_{B,k}')}{q(\tau_{B,k}'\mid\tau_{B,k})}
\right\},
\]

\Statex
\State \text{Nugget updates} ($\eta_{k,m}$ for $m=1,\dots,D$): for each $m$, propose $\log\eta_{k,m}'=\log\eta_{k,m}+\epsilon_{\eta,m}$, $\epsilon_{\eta,m}\sim\mathcal{N}(0,s_\eta^2)$

\State \hspace{1.2em}Acceptance:
\[
\alpha_{\eta,m}=\min\!\left\{1,\;
\exp(\ell'-\ell)\,
\frac{p(\eta_{k,m}')}{p(\eta_{k,m})}\,
\frac{q(\eta_{k,m}\mid\eta_{k,m}')}{q(\eta_{k,m}'\mid\eta_{k,m})}
\right\}.
\]
\Statex
\State \text{Shape update} ($B_k^{\text{shape}}$):
\State \hspace{1.2em}maintain lower-triangular $L$ with $B_k^{\text{shape}}=D\,\big(L L^\top\big)/\mathrm{tr}(L L^\top)$
\State \hspace{1.2em}propose $L' = L + E$ where $E_{ab}\sim\mathcal{N}(0,s_B^2)$ for $a\ge b$ and $E_{ab}=0$ for $a<b$ 
\State \hspace{1.2em}set $S'=L'L'^\top$ and $B_k^{\text{shape}\,'}=D\,S'/\mathrm{tr}(S')$
\State \hspace{1.2em}accept with $\alpha_B=\min\{1,\exp(\ell'-\ell)\,p(B_k^{\text{shape}\,'})/p(B_k^{\text{shape}})\}$.
\end{algorithmic}
\end{algorithm}

\section{Simulation Study}
\label{chap:simstudy}

This section presents a Monte Carlo simulation study evaluating the proposed Bayesian nonparametric model, WICMAD, for detecting anomalous functional data. We compare its performance against two benchmark approaches implemented in the DDAlpha and MRFDepth R packages. Specifically, we look at the band depth \citep{lopez2009concept} and functional
adjusted outlyingness \citep{hubert_multivariate_2015}. Both competing methods output anomaly scores that must be thresholded to obtain class labels. For these methods, class labels are assigned using the cutoff that maximizes the F1-score. It should be noted, however, that this represents a best-case scenario, as real applications would require a selection criterion to obtain an appropriate cutoff. Section~\ref{sec:sim-design} presents the simulated data and Section~\ref{sec:sim-results} showcases model performance and provides a discussion.

\subsection{Simulation Design}
\label{sec:sim-design}

All data are generated from a Gaussian process evaluated on an evenly spaced grid of $P=32$ time points over $\mathcal{T}=[0,1]$. Each function is constructed by combining a structured mean function with a quasiperiodic covariance kernel and additive Gaussian noise $\mathcal{N}(0,0.05^2)$.

The mean function captures both periodic and trend components and is defined as
\[
m(t) = 0.6\sin(4\pi t) + 0.25\cos(10\pi t) + 0.1t.
\]
Dependence within the function is generated by a quasiperiodic squared exponential kernel,
\[
k_{\text{QP}}(t,t')
=\sigma_1^2
\exp\!\Big(-\frac{(t-t')^2}{2\ell_1^2}\Big)
\exp\!\Big(-\frac{2\sin^2\!\big(\pi(t-t')/\omega\big)}{\ell_\omega^2}\Big)
+\sigma_2^2
\exp\!\Big(-\frac{(t-t')^2}{2\ell_2^2}\Big),
\]
which combines a periodic component with a smooth long-range trend. The kernel parameters are set to $(\ell_1,\sigma_1^2,\omega,\ell_\omega)=(0.15,1.0,0.30,0.30)$ for the quasiperiodic component and $(\ell_2,\sigma_2^2)=(0.60,0.4)$ for the squared exponential term.

15\% of the $N=300$ curves are made anomalous by adding one of four perturbations. For our anomaly types, we mimic the work of \citet{staerman_functional_2023}. The isolated anomaly introduces a narrow Gaussian bump at a random location:
\begin{align*}
a_i^{\text{iso}}(t)
&= s\,a\,\exp\!\Big(-\frac{(t-t_{i_0})^2}{2w^2}\Big).
\end{align*}
Here, $s\in\{-1,+1\}$ is a random sign, $a\sim\operatorname{Unif}(8,12)$ is the amplitude, and $w\sim\operatorname{Unif}(0.3,0.8)$ is the width of the bump. The center index $i_0$ is drawn uniformly from $\{3,\ldots,P-2\}$. 

The magnitude I anomaly applies a constant level shift across the entire domain:
\begin{align*}
a_i^{\text{mag1}}(t)
&= s\,a.
\end{align*}
The shift direction is determined by $s\in\{-1,+1\}$, and the amplitude is sampled as $a\sim\operatorname{Unif}(12,15)$.

The magnitude II anomaly introduces a localized bump on a short subinterval:
\begin{align*}
a_i^{\text{mag2}}(t_j)
&=
\begin{cases}
\dfrac{s\,a}{2}\!\left(1-\cos\!\Big(\pi\dfrac{j-j_0}{\lambda-1}\Big)\right), & j\in\{s,\ldots,s+\lambda-1\},\\
0, & \text{otherwise.}
\end{cases}
\end{align*}
The window length is $\lambda=\lfloor0.10\,P\rfloor$, and the start index $j_0$ is drawn uniformly from $\{1,\ldots,P-\lambda+1\}$. As before, $s\in\{-1,+1\}$ selects the direction and $a\sim\operatorname{Unif}(10,15)$ controls the amplitude. 

The shape anomaly adds a sinusoid with random frequency:
\begin{align*}
a_i^{\text{shape}}(t)
&= 3\,\sin(2\pi U t),
\end{align*}
where $U$ is drawn from $\operatorname{Unif}(0.2,2.0)$. 

Multivariate datasets with $D=3$ dimensions are generated by mixing two latent Gaussian process factors through a fixed loading matrix
\[
A = \begin{bmatrix}
    1.0 & 0.4 \\
    0.2 & 1.0 \\
    0.7 & -0.3
\end{bmatrix}
\]
which gives rise to curves
\[
X_i(t) = U_i(t)A^\top + E_i(t),
\qquad
U_i(t)=\big(u_{i1}(t),\,u_{i2}(t)\big),\quad
u_{iq}(t)\sim\operatorname{GP}\!\big(m(t),\,k_{\text{QP}}(t,t')\big),
\]
where $E_i(t)\sim \mathcal{N}(0,\sigma^2 I_3)$ adds independent noise.  
Anomalies are added per channel using the same approach as above.

We test three multivariate datasets where anomalies of any type can appear and the number of dimensions that contain anomalies increases with each dataset. Each dataset is analyzed under a semi-supervised setting where 15\% of normal curves are revealed to the model. For every dataset presented in this section, we then ran four Markov chains of length 55,000. After discarding an initial burn-in of 5,000 iterations and applying a thinning factor of 10, we retained 5,000 posterior samples from each chain. Running multiple chains is necessary due to the highly multimodal nature of the posterior distribution and allows us to assess convergence across chains. In addition to standard convergence diagnostics, we also monitor log-likelihood trajectories, as the sampler may become trapped in local modes.

\subsection{Results}
\label{sec:sim-results}

\begin{figure}[ht]
  \centering
  \begin{subfigure}[t]{0.48\linewidth}
    \centering
    \includegraphics[width=\linewidth]{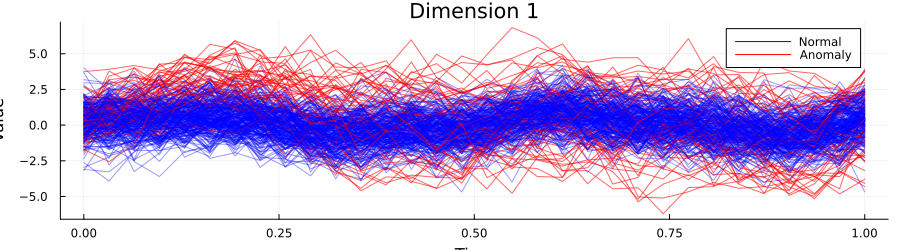}
    \caption{Shape anomaly dataset.}
    \label{fig:other:shape}
  \end{subfigure}\hfill
  \begin{subfigure}[t]{0.48\linewidth}
    \centering
    \includegraphics[width=\linewidth]{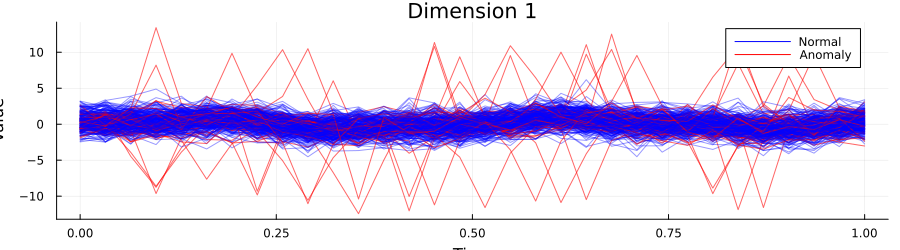}
    \caption{Isolated anomaly dataset.}
    \label{fig:other:isolated}
  \end{subfigure}

  \vspace{0.75em}

  \begin{subfigure}[t]{0.48\linewidth}
    \centering
    \includegraphics[width=\linewidth]{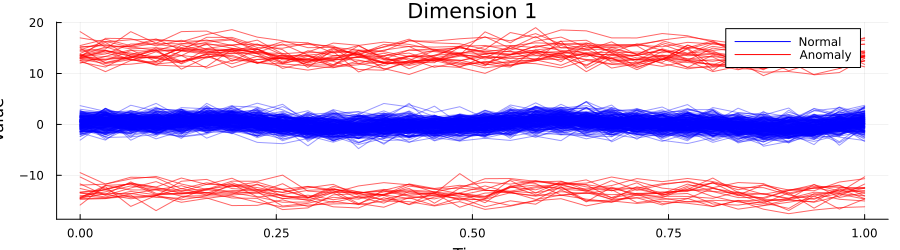}
    \caption{Magnitude 1 anomaly dataset.}
    \label{fig:other:mag1}
  \end{subfigure}\hfill
  \begin{subfigure}[t]{0.48\linewidth}
    \centering
    \includegraphics[width=\linewidth]{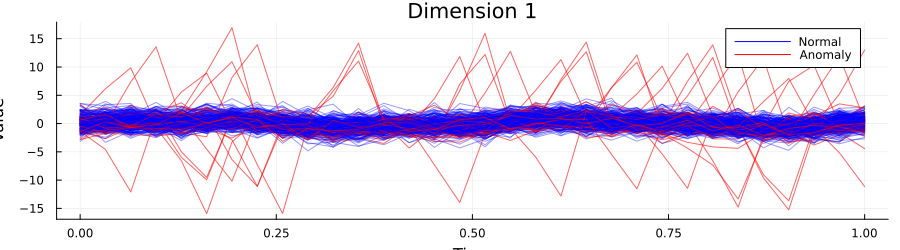}
    \caption{Magnitude 2 anomaly dataset.}
    \label{fig:other:mag2}
  \end{subfigure}
  \caption{Simulated univariate functional data each with a different anomaly type.}
  \label{fig:others}
\end{figure}

\begin{figure}[ht]
  \centering
  \includegraphics[width=0.8\linewidth]{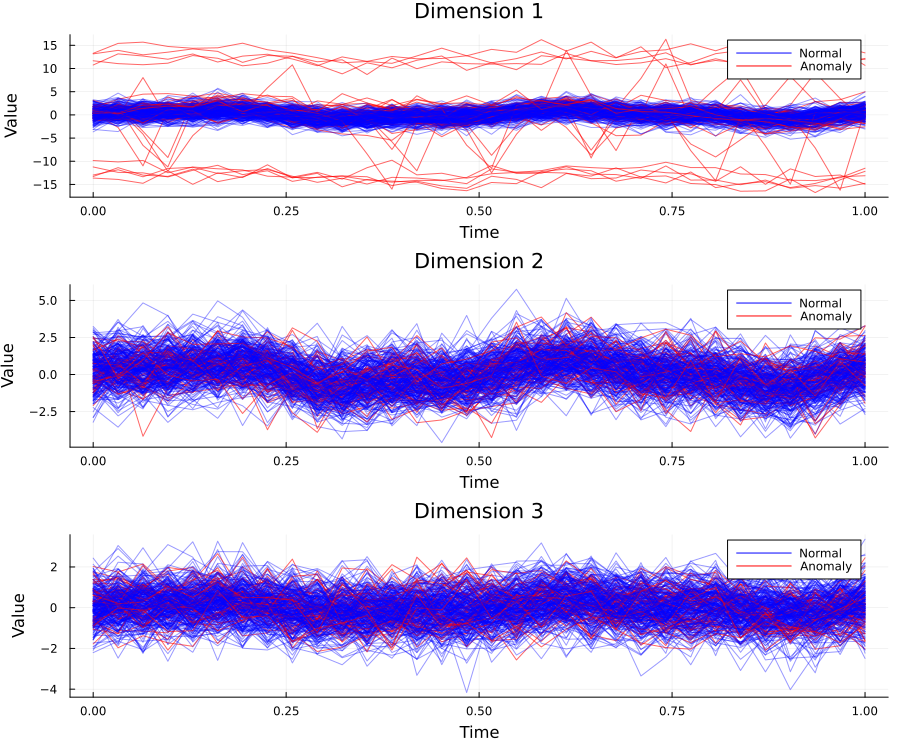}
  \caption{Simulated multivariate functional data with anomalies in one dimension.}
  \label{fig:channels:one}
\end{figure}

\begin{figure}[ht]
  \centering
  \includegraphics[width=0.8\linewidth]{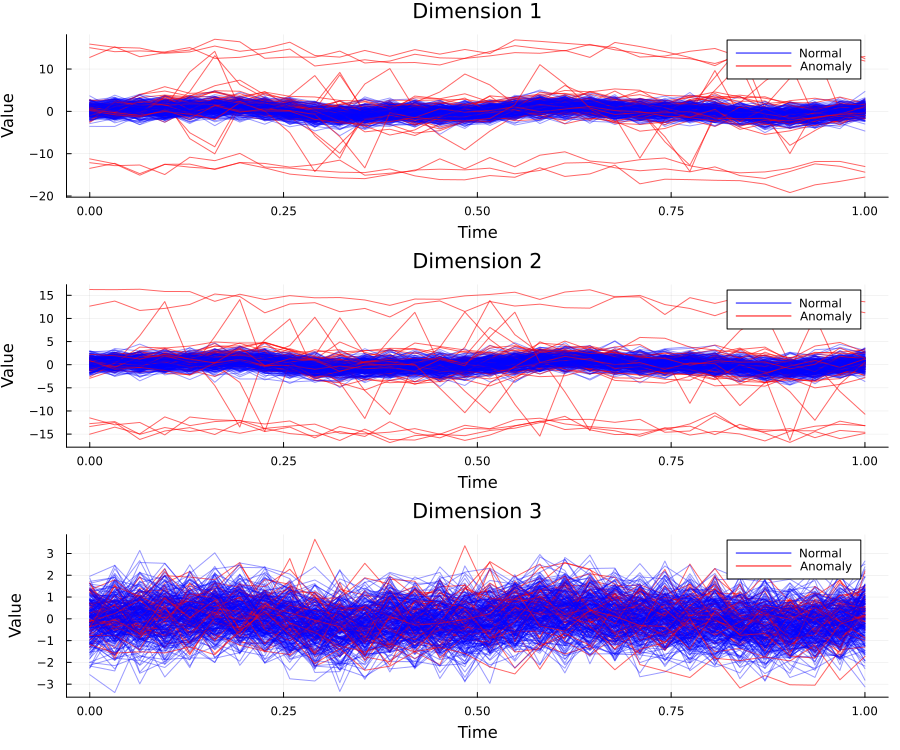}
  \caption{Simulated multivariate functional data with anomalies in two dimensions.}
  \label{fig:channels:two}
\end{figure}

\begin{figure}[ht]
  \centering
  \includegraphics[width=0.8\linewidth]{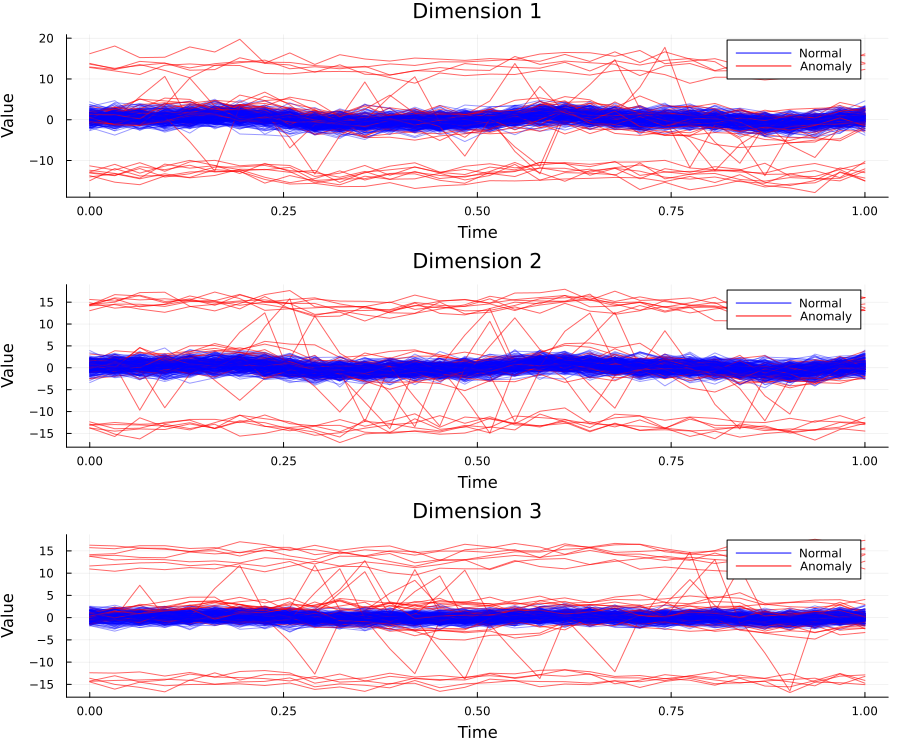}
  \caption{Simulated multivariate functional data with anomalies in all three dimensions.}
  \label{fig:channels:three}
\end{figure}

This section presents results across the seven simulated datasets. For each dataset, we report median accuracy, precision, recall, and F1-score across 100 Monte Carlo replications. Results for univariate and multivariate settings are summarized in Tables~\ref{tab:sim-summary-uni} and \ref{tab:sim-summary-multi}, respectively. Visualizations of the simulated datasets are provided in Figures~\ref{fig:others}, \ref{fig:channels:one}, \ref{fig:channels:two}, and \ref{fig:channels:three}.

\begin{table}[t]
    \centering
    \caption{Median performance metrics across 100 Monte Carlo replications for the univariate datasets.}
    \label{tab:sim-summary-uni}
    \begin{tabular}{llcccc}
        \toprule
        Dataset & Method & Accuracy & Precision & Recall & F1-score \\
        \midrule

        \textbf{Univariate: Isolated}
        & WICMAD   & \textbf{0.99} & \textbf{1.00} & 0.96 & \textbf{0.99} \\
        & DDAlpha  & 0.15 & 0.15 & \textbf{1.00} & 0.26 \\
        & MRFDepth & 0.86 & 0.62 & 0.22 & 0.33 \\
        \midrule

        \textbf{Univariate: Magnitude I}
        & WICMAD   & \textbf{1.00} & \textbf{1.00} & \textbf{1.00} & \textbf{1.00} \\
        & DDAlpha  & 0.95 & \textbf{1.00} & 0.69 & 0.82 \\
        & MRFDepth & 0.90 & \textbf{1.00} & 0.36 & 0.52 \\
        \midrule

        \textbf{Univariate: Magnitude II}
        & WICMAD   & \textbf{1.00} & \textbf{1.00} & \textbf{1.00} & \textbf{1.00} \\
        & DDAlpha  & 0.15 & 0.15 & \textbf{1.00} & 0.26 \\
        & MRFDepth & 0.88 & 0.81 & 0.29 & 0.43 \\
        \midrule

        \textbf{Univariate: Shape}
        & WICMAD   & \textbf{0.97} & \textbf{1.00} & 0.82 & \textbf{0.90} \\
        & DDAlpha  & 0.16 & 0.15 & \textbf{1.00} & 0.26 \\
        & MRFDepth & 0.90 & \textbf{1.00} & 0.36 & 0.52 \\

        \bottomrule
    \end{tabular}
\end{table}

\begin{table}[t]
    \centering
    \caption{Median performance metrics across 100 Monte Carlo replications for the multivariate datasets.}
    \label{tab:sim-summary-multi}
    \begin{tabular}{llcccc}
        \toprule
        Dataset & Method & Accuracy & Precision & Recall & F1-score \\
        \midrule

        \textbf{Multivariate: One Dimension}
        & WICMAD   & \textbf{1.00} & \textbf{1.00} & \textbf{1.00} & \textbf{1.00} \\
        & DDAlpha  & 0.72 & 0.31 & 0.67 & 0.45 \\
        & MRFDepth & 0.90 & \textbf{1.00} & 0.36 & 0.52 \\
        \midrule

        \textbf{Multivariate: Two Dimensions}
        & WICMAD   & \textbf{1.00} & \textbf{1.00} & \textbf{0.98} & \textbf{1.00} \\
        & DDAlpha  & 0.72 & 0.30 & 0.69 & 0.43 \\
        & MRFDepth & 0.90 & \textbf{1.00} & 0.36 & 0.52 \\
        \midrule

        \textbf{Multivariate: Three Dimensions}
        & WICMAD   & \textbf{1.00} & \textbf{1.00} & \textbf{1.00} & \textbf{1.00} \\
        & DDAlpha  & 0.71 & 0.32 & 0.69 & 0.46 \\
        & MRFDepth & 0.90 & \textbf{1.00} & 0.36 & 0.52 \\

        \bottomrule
    \end{tabular}
\end{table}

The simulation study demonstrates that the proposed model can detect a range of functional anomalies, including changes in magnitude, local behavior, and overall shape. The wavelet representation captures localized departures from the baseline signal, while kernel switching accommodates differences in global dependence structure across clusters. We first discuss the univariate results before turning to the multivariate setting.

Across all univariate experiments, the model achieves near-perfect median performance in terms of accuracy, precision, recall, and F1-score. As shown in Table~\ref{tab:sim-summary-uni}, WICMAD remains competitive across all datasets and outperforms the benchmark methods in several cases, particularly for the Magnitude II setting. Notably, WICMAD consistently attains high F1-scores, indicating a strong balance between detecting anomalous curves and limiting false positives. It achieves the best F1 performance on the Isolated and Magnitude II datasets and performs comparably well on the Magnitude I and Shape datasets.

The multivariate results in Table~\ref{tab:sim-summary-multi} show a similar pattern. WICMAD achieves perfect median accuracy, precision, recall, and F1-score when anomalies affect one or three dimensions, and remains near-perfect when anomalies affect two dimensions, with median recall of 0.98 and F1-score of 1.00. This indicates that the model is able to use cross-dimensional dependence to identify anomalous curves even when departures are restricted to only a subset of the functional coordinates. In contrast, DDAlpha attains moderate recall but low precision in the multivariate settings, while MRFDepth achieves perfect precision but substantially lower recall. Thus, the competing methods tend either to over-detect anomalies or miss a large fraction of them, whereas WICMAD maintains a more favorable balance between false positives and false negatives.

It should be noted that we report median performance across Monte Carlo replications due to occasional degeneracy in the clustering structure, where the model collapses to a single cluster. Such behavior can arise when the model is overly flexible, allowing variance components to inflate and absorb anomalous observations into the baseline cluster. In practice, this flexibility is primarily determined by the choice of covariance kernels included in the Carlin--Chib selection step. As a result, careful specification of the kernel library is important. For this reason, we recommend applying the model in semi-supervised or supervised settings, where candidate kernels can be validated.

\section{Real-Data Applications}
\label{chap:apps}

This section reports results from applying WICMAD to three real datasets: Character Trajectories \citep{character_trajectories_175} in Section~\ref{sec:ct}, Asphalt Regularity \citep{souza2018asphalt} in Section~\ref{sec:ar}, and Chinatown \citep{Chinatown_TSC_dataset} in Section~\ref{sec:china}. We conclude with Section~\ref{sec:appdis}, which offers a detailed discussion of the results.

For each dataset, we first interpolated the data to a dimensionality of 32. For every analysis presented in this section, we then ran four Markov chains of length 55,000. After discarding an initial burn-in of 5,000 iterations and applying a thinning factor of 10, we retained 5,000 posterior samples from each chain. Running multiple chains is necessary due to the highly multimodal nature of the posterior distribution and allows us to assess convergence across chains. In addition to standard convergence diagnostics, we also monitor log-likelihood trajectories, as the sampler may become trapped in local modes. For more details see Appendix~\ref{app:con}

We present the confusion matrix, precision, recall, and F1 scores for each dataset tested. Similar to Section~\ref{chap:simstudy}, each dataset was artificially imbalanced to reflect the anomaly detection setting. That is, we selected one group to be the normal class, included all observations, and then artificially imbalanced the dataset to have 15\% from another class. We also assume a semi-supervised setting and revealed 15\% of the normal group's labels to the model. 

Just as in Section~\ref{chap:simstudy}, we compare our model’s performance against two benchmark approaches implemented in the DDAlpha and MRFDepth R packages. Specifically, we look at the band depth \citep{lopez2009concept} and functional
adjusted outlyingness \citep{hubert_multivariate_2015}. For these methods, class labels are assigned using the cutoff that maximizes the F1-score. It should be noted, however, that this represents a best-case scenario, as real applications would require a selection criterion to obtain an appropriate cutoff.

\subsection{Character Trajectories}
\label{sec:ct}


The Character Trajectories dataset is a multivariate functional dataset consisting of pen tip trajectories recorded while a single writer produced handwritten characters. The data were originally collected for the purpose of primitive extraction and contain multiple labeled realizations of individual characters. Only characters written using a single continuous pen-down stroke were retained in the dataset.

Each observation is a three-dimensional functional curve recorded over time, corresponding to the horizontal pen position, vertical pen position, and pen-tip force applied during writing. The trajectories were sampled at a frequency of 200 Hz and subsequently smoothed during preprocessing. The dataset contains 2858 labeled samples distributed across multiple character classes corresponding to letters of the alphabet. Additional details on data acquisition and preprocessing are provided in \citet{character_trajectories_175}. 

\begin{figure}
    \centering
\includegraphics[width=1\linewidth]{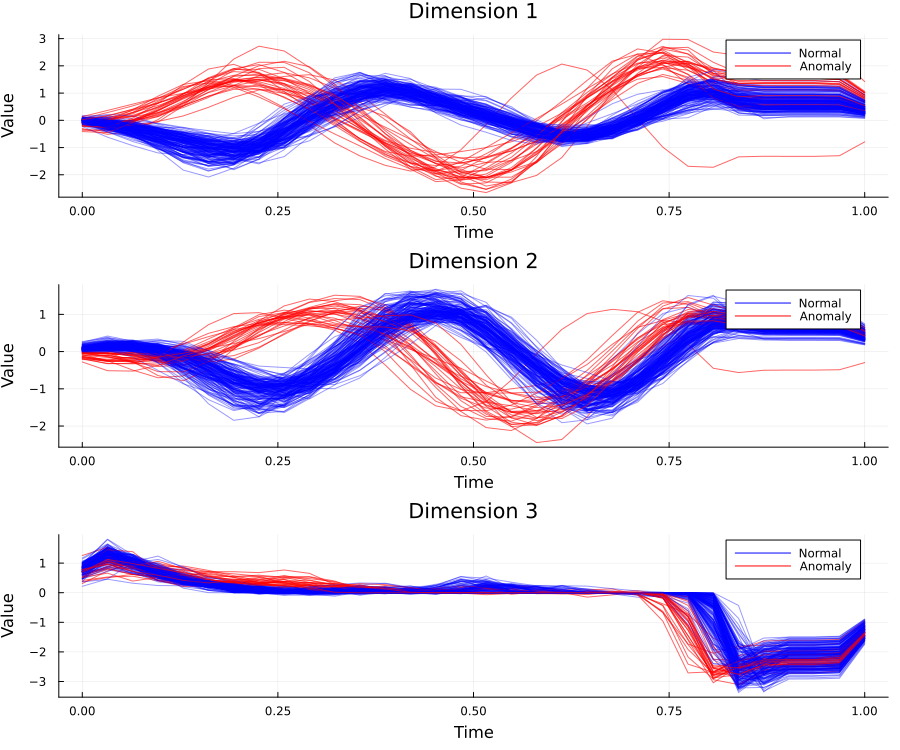}
    \caption{The Character Trajectories dataset colored by normal (blue) and anomalous (red) class labels. }
    \label{fig:characters}
\end{figure}

For the anomaly detection experiment, we construct a binary classification setting by designating the letter ``a'' as the normal class and the letter ``b'' as the anomalous class, discarding all other character classes. This yields a two-group dataset in which the normal trajectories represent repeated realizations of a common writing motion, while anomalies correspond to structurally distinct pen trajectories arising from a different character. Figure~\ref{fig:characters} illustrates representative trajectories colored by class label. Quantitative results for this dataset are reported in Tables~\ref{tab:char1} and~\ref{tab:char2}.

\begin{table}[ht]
\centering

\begin{minipage}[t]{0.55\linewidth}
\centering
\begin{tabular}{clcc}
 & & \multicolumn{2}{c}{Predicted} \\
 & & Normal & Anomaly \\
\multirow{2}{*}{True} & Normal  & 170 & 0 \\
                      & Anomaly & 0   & 30 \\
\end{tabular}
\captionof{table}{Maximum a posteriori confusion matrix for the Character Trajectories dataset using the WICMAD method.}
\label{tab:char1}
\end{minipage}
\hfill
\begin{minipage}[t]{0.38\linewidth}
\centering
\begin{tabular}{lccc}
\toprule
Method   & Precision & Recall & F1 \\
\midrule
WICMAD   & \textbf{1.00} & \textbf{1.00} & \textbf{1.00} \\
MRFDepth & 0.47          & 0.90          & 0.61          \\
DDAlpha  & 0.15          & \textbf{1.00} & 0.26          \\
\bottomrule
\end{tabular}
\captionof{table}{Performance metrics for the Character Trajectories dataset.}
\label{tab:char2}
\end{minipage}

\end{table}

\subsection{Asphalt Regularity}
\label{sec:ar}


Next, we consider the Asphalt Regularity dataset, which consists of accelerometer measurements collected from a smartphone mounted inside a vehicle while driving over roads with different pavement conditions. The data were collected using a flexible suction mount positioned near the dashboard, while an Android application called \emph{Asfault} continuously recorded sensor measurements and pavement labels \citep{souza2018asphalt}. During data collection, an expert driver labeled the pavement condition in real time, and video recordings of the road surface were captured to verify label accuracy.

The smartphone accelerometer records acceleration along three physical axes, producing time series corresponding to the x, y, and z acceleration components. These signals were sampled at approximately 100 Hz. Because Android sensor acquisition does not guarantee perfectly uniform timing, the effective sampling rate varies slightly. In addition to acceleration, the system records GPS location and velocity, though only the accelerometer measurements are used in this work. To obtain a functional representation, the three-axis acceleration signals are combined into a single acceleration magnitude time series resulting in a univariate functional dataset.

\begin{figure}
    \centering
\includegraphics[width=1\linewidth]{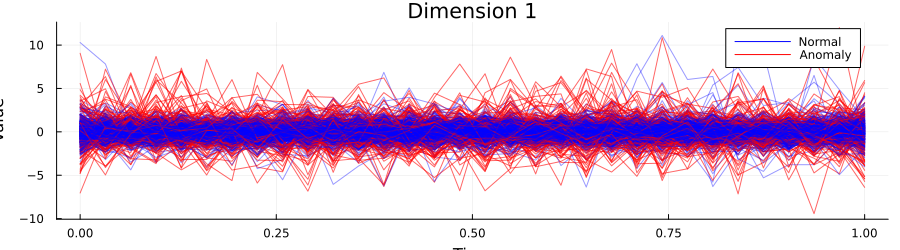}
    \caption{The Asphalt Regularity dataset colored by normal (blue) and anomalous (red) class labels. }
    \label{fig:asphalt}
\end{figure}

Regular pavement trajectories are treated as the normal class, and the deteriorated pavement trajectories are treated as anomalies. Figure~\ref{fig:asphalt} displays example trajectories colored by class label. Results for this dataset are reported in Tables~\ref{tab:as1} and~\ref{tab:as2}.

\begin{table}[ht]
\centering

\begin{minipage}[t]{0.55\linewidth}
\centering
\begin{tabular}{clcc}
 & & \multicolumn{2}{c}{Predicted} \\
 & & Normal & Anomaly \\
\multirow{2}{*}{True} & Normal  & 685 & 74 \\
                      & Anomaly & 5   & 129 \\
\end{tabular}
\captionof{table}{Maximum a posteriori confusion matrix for the Asphalt Regularity dataset using the WICMAD method.}
\label{tab:as1}
\end{minipage}
\hfill
\begin{minipage}[t]{0.38\linewidth}
\centering
\begin{tabular}{lccc}
\toprule
Method   & Precision & Recall & F1 \\
\midrule
WICMAD   & 0.64          & 0.96          & 0.77 \\
MRFDepth & \textbf{0.79} & 0.88          & \textbf{0.83} \\
DDAlpha  & 0.15          & \textbf{1.00} & 0.26 \\
\bottomrule
\end{tabular}
\captionof{table}{Performance metrics for the Asphalt Regularity dataset.}
\label{tab:as2}
\end{minipage}

\end{table}

\subsection{Chinatown}
\label{sec:china}


The final dataset we consider is the Chinatown dataset, which consists of pedestrian count time series collected at an urban intersection in Melbourne, Australia. The data come from the City of Melbourne’s automated pedestrian counting system, which is designed to monitor pedestrian activity patterns across the municipality and support urban planning and infrastructure decisions. The specific series used in this study corresponds to pedestrian counts recorded at the Chinatown--Swanston Street (North) location over the 2017 calendar year \citep{Chinatown_TSC_dataset}.

Each observation is a univariate time series representing pedestrian volume measured over the course of a day. The classification task associated with this dataset is to distinguish between weekday and weekend pedestrian activity patterns. In this work, weekday observations are treated as the normal class while weekend observations are treated as anomalies.

\begin{figure}
    \centering
\includegraphics[width=1\linewidth]{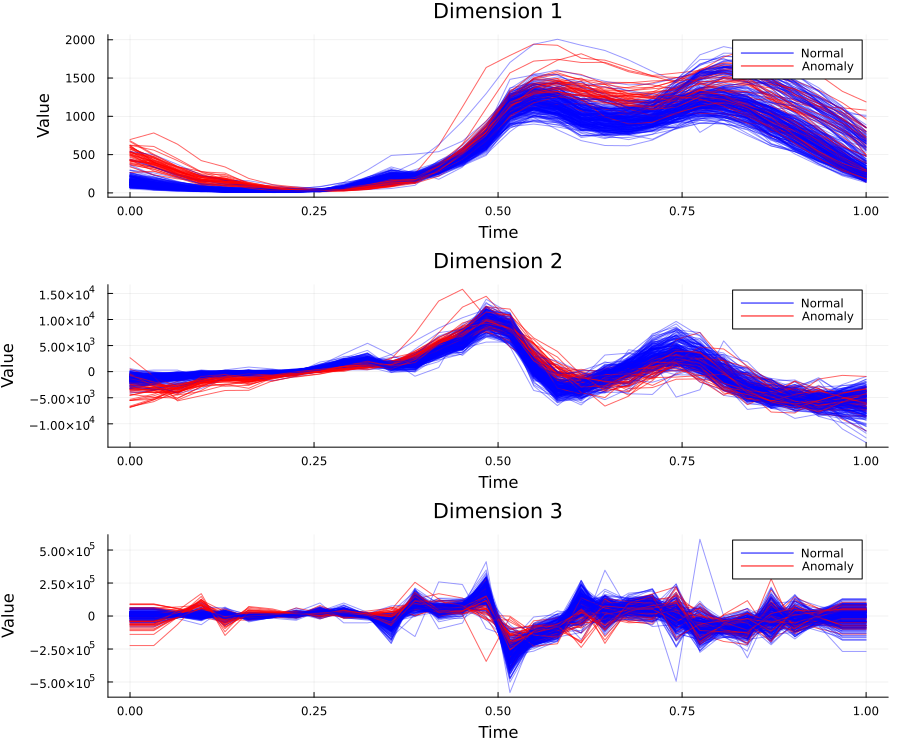}
    \caption{The Chinatown dataset colored by normal (blue) and anomalous (red) class labels. The data are augmented with the first and second derivatives estimated using finite differencing.}
    \label{fig:china}
\end{figure}

To improve separability between the two groups, we augment each trajectory with an estimate of its first and second derivatives computed using finite differencing. This results in a three-dimensional functional representation consisting of pedestrian counts and their estimated rate of change and acceleration over time. Figure~\ref{fig:china} displays the trajectories colored according to class membership. Results for this dataset are reported in Tables~\ref{tab:china1} and~\ref{tab:china2}.

\begin{table}[ht]
\centering

\begin{minipage}[t]{0.55\linewidth}
\centering
\begin{tabular}{clcc}
 & & \multicolumn{2}{c}{Predicted} \\
 & & Normal & Anomaly \\
\multirow{2}{*}{True} & Normal  & 249 & 6 \\
                      & Anomaly & 1   & 44 \\
\end{tabular}
\captionof{table}{Maximum a posteriori confusion matrix for the Chinatown dataset using the WICMAD method.}
\label{tab:china1}
\end{minipage}
\hfill
\begin{minipage}[t]{0.38\linewidth}
\centering
\begin{tabular}{lccc}
\toprule
Method   & Precision & Recall & F1 \\
\midrule
WICMAD   & \textbf{0.88} & 0.98 & \textbf{0.93} \\
MRFDepth & 0.33 & 0.91 & 0.48 \\
DDAlpha  & 0.17 & \textbf{1.00} & 0.29 \\
\bottomrule
\end{tabular}
\captionof{table}{Performance metrics for the Chinatown dataset.}
\label{tab:china2}
\end{minipage}

\end{table}

\subsection{Discussion}
\label{sec:appdis}

This section demonstrates that the proposed model generalizes well to real-world data, achieving strong performance across datasets with varying levels of smoothness and structural complexity. Overall, the model consistently attains high recall, often at the expense of some precision, resulting in competitive F1-scores in the anomaly detection setting.

The Character Trajectories dataset represents an ideal setting for the model. These data consist of smooth, structured pen trajectories that are well aligned with the Gaussian process assumption. As shown in Tables~\ref{tab:char1} and~\ref{tab:char2}, the model achieves perfect classification performance across all metrics. This is consistent with the simulation results, where shape-based anomalies were detected reliably. In contrast, the competing methods exhibit substantially lower precision and F1-scores despite achieving high recall, indicating a tendency to over-classify anomalies.

The Asphalt Regularity dataset presents a more challenging scenario. As seen in Tables~\ref{tab:as1} and~\ref{tab:as2}, the model achieves a high recall of 0.96 but a lower precision of 0.64, resulting in an F1-score of 0.77. These data are less smooth and exhibit greater variability than the Character Trajectories dataset, making them less well suited to the Gaussian process assumption. Despite this, the model successfully identifies nearly all anomalous observations, with the reduction in precision driven by an increase in false positives. While the MRFDepth method attains a higher F1-score of 0.83, it does so with a lower recall, indicating that it misses more anomalies. In many anomaly detection applications, this trade-off is undesirable, as failing to detect anomalies is typically more costly than flagging false positives.

The Chinatown dataset was selected to further assess performance on nonstationary and less structured data. As shown in Tables~\ref{tab:china1} and~\ref{tab:china2}, the model achieves strong performance with a precision of 0.88, recall of 0.98, and an F1-score of 0.93. The confusion matrix indicates that nearly all anomalous observations are correctly identified, with 44 out of 45 detected and only a single false negative, while maintaining a low number of false positives at 6. This suggests that the model is able to effectively distinguish weekday and weekend pedestrian patterns despite their inherent variability. In contrast, DDAlpha attains perfect recall of 1.00 but very low precision of 0.17, indicating that it effectively classifies most observations as anomalous, while MRFDepth achieves high recall of 0.91 but low precision of 0.33, resulting in a substantially lower F1-score of 0.48.

Overall, these experiments demonstrate that the model performs well across a range of real-world settings, including cases where the Gaussian process assumption is only approximately satisfied. The results highlight a consistent pattern in which the model prioritizes high recall, making it well suited for applications where detecting anomalous behavior is critical. At the same time, performance is sensitive to the choice of covariance kernels and model specification. Expanding the kernel set or incorporating additional mechanisms to account for nonstationarity may further improve performance, particularly in more challenging datasets.

\section{Conclusion}
\label{sec:conclu}

This article introduced WICMAD, a Bayesian nonparametric model for semi-supervised anomaly detection in multivariate functional data. The model represents functional observations using an infinite mixture of multi-output Gaussian processes, with wavelet-based mean functions and Besov priors used to capture sparse local structure. Dependence across functional dimensions is modeled through an intrinsic coregionalization model, while a Carlin--Chib product-space step allows covariance kernels to be selected in a cluster-specific manner during MCMC inference.

The simulation study showed that WICMAD can detect a range of functional anomaly types in both univariate and multivariate settings. Across the simulated datasets, the method achieved strong median accuracy, precision, recall, and F1-scores, including cases where anomalies affected only a subset of the functional dimensions. These results suggest that combining wavelet-based mean modeling with multivariate Gaussian process dependence provides a useful mechanism for separating anomalous curves from the dominant data-generating structure.

The real-data applications demonstrated that WICMAD performs well across a range of functional data settings. The method achieved perfect classification on the Character Trajectories dataset, strong performance on the Chinatown dataset, and high recall on the more challenging Asphalt Regularity dataset. These datasets span smooth handwritten trajectories, high-variance accelerometer signals, and nonstationary pedestrian count data, suggesting that the approach remains robust under moderate departures from the Gaussian process assumption. A consistent finding across all applications was the model's tendency to prioritize recall over precision, making it particularly attractive in settings where missed anomalies are more costly than false alarms.

Several directions remain for future work. The Gaussian process formulation naturally supports missing-data and imputation problems, which were not explored here. It may also be useful to expand the kernel library, consider kernel combinations, or develop covariance structures better suited to nonstationary variance. Finally, alternative stochastic processes such as Student-$t$ processes, as well as alternatives to the Dirichlet process such as Pitman--Yor or dependent Dirichlet processes, may provide additional flexibility for modeling outliers and covariate-dependent clustering structure.


\acks{This work was supported by Fonds de recherche du Québec -- Nature et technologies (FRQNT). The authors have no competing interests.}


\newpage

\appendix
\section{Full Hierarchical Model}
\label{sec:full-model}

We now present the full hierarchical specification.  
Let $\theta_k$ denote the complete set of cluster--specific parameters,
\[
\theta_k
=
\Big(
\{\beta_{k,m,j,\ell}\},\;
\{\gamma_{k,m,j,\ell}\},\;
\{g_{k,j}\},\;
\{\pi_{k,j}\},\;
\{\sigma_{k,m}^2\},\;
\tau_{\sigma,k},\;
\tau_{B,k},\;
B_k^{\text{shape}},\;
\eta_k,\;
s_k,\;
\{\phi_{k,r}\}_{r=1}^R
\Big).
\]

A Dirichlet process prior is placed on the unknown distribution of cluster
parameters. Let curve $Y_i$ be in cluster $k$, then the generative model is given by
\begin{align*}
    Y_i \mid \theta_{k} &\sim F(\theta_{k}), \\
    \theta_k \mid G &\stackrel{iid}{\sim} G, \\
    G &\sim \mathrm{DP}(\xi, G_0),
\end{align*}
where $F(\theta)$ denotes the sampling distribution of a curve, $\xi$ is the concentration parameter, and $G_0$ is the base measure. Using the stick-breaking representation of the Dirichlet process \citep{sethuraman1994constructive}, the resulting hierarchical model becomes:
\begin{align*}
\xi &\sim \mathrm{Gamma}(a_\xi,b_\xi)
\\[6pt]
v_k \mid \xi &\sim \mathrm{Beta}(1,\xi),
\qquad
w_k = v_k \prod_{j<k} (1 - v_j)
\\[6pt]
z_i \mid \{w_k\} &\sim \mathrm{Categorical}(\{w_k\})
\\[12pt]
\pi_{k,j} &\sim \mathrm{Beta}(\tau_\pi m_j,\;\tau_\pi (1 - m_j)),
\qquad
m_j = \kappa_\pi\,2^{-c_2 j}
\\[6pt]
\gamma_{k,m,j,\ell} \mid \pi_{k,j}
&\sim \mathrm{Bernoulli}(\pi_{k,j})
\\[6pt]
\beta_{k,m,j,\ell} \mid \gamma_{k,m,j,\ell}, g_{k,j}, \sigma_{k,m}^2
&\sim (1-\gamma_{k,m,j,\ell})\,\delta_0
   + \gamma_{k,m,j,\ell}\,\mathcal{N}(0,\,g_{k,j}\sigma_{k,m}^2)
\\[6pt]
g_{k,j} &\sim \mathrm{Inv\text{-}Gamma}(a_g,b_g)
\\[6pt]
\sigma_{k,m}^2 \mid \tau_{\sigma,k}
&\sim \mathrm{Inv\text{-}Gamma}(a_\sigma,\;b_\sigma\,\tau_{\sigma,k})
\\[6pt]
\tau_{\sigma,k} &\sim \mathrm{Gamma}(a_\tau,b_\tau)
\\[12pt]
B_k^{\text{shape}} &= D\,\dfrac{L_k L_k^\top}{\mathrm{tr}(L_k L_k^\top)}
\\[4pt]
(L_k)_{ab} &\sim \mathcal{N}(0,1), \qquad a\ge b
\\[6pt]
\eta_{k,m} &\sim \mathrm{Inv\text{-}Gamma}(a_\eta,b_\eta)
\\[12pt]
\phi_{k,r} &\sim p_r(\cdot), \qquad r=1,\dots,R
\\[4pt]
s_k &\sim \mathrm{Categorical}\!\left(\tfrac{1}{R},\dots,\tfrac{1}{R}\right)
\\[12pt]
\mathrm{vec}(Y_i - \mu_{z_i}) \mid z_i=k,\theta_k
&\sim 
\mathcal{N}\!\Big(
0,\;
K_x^{(s_k)}(\phi_{k,s_k}) \otimes \big(\tau_{B,k}\,B_k^{\text{shape}}\big)
   + I_P \otimes \mathrm{diag}(\eta_k)
\Big).
\end{align*}

\section{MCMC Updates}
\label{app:post}
\subsection{Wavelet Posterior Updates}
\label{sec:wavelet_post}
In this section, the posterior updates for the wavelet model are derived under the
assumption that curves assigned to the same cluster share a mean latent wavelet coefficient for each location and scale. We further assume each observed coefficient for a curve is a noisy draw around this shared mean. For a given wavelet coefficient we observe values $y_1,\ldots,y_n$ from $n$ curves belonging to the same cluster. These observations follow

\[
y_i = \beta + \varepsilon_i, \qquad \varepsilon_i \sim \mathcal N(0,\sigma^2), \quad i=1,\ldots,n,
\]

where the coefficient $\beta$ is shared across all curves in the cluster and the noise terms are independent with variance $\sigma^2$. The spike and slab prior introduces the latent inclusion indicator $\gamma \in \{0,1\}$ such that

\[
\beta \mid \gamma,g,\sigma^2 \sim (1-\gamma)\delta_0 + \gamma\,\mathcal N(0,\sigma^2 g),
\qquad
\Pr(\gamma = 1) = \pi.
\]
\noindent Define $S_1 = \sum_{i=1}^n y_i$ and the sample mean $\bar y = (1/n)\sum_{i=1}^n y_i$. If $\gamma = 0$ then $\beta$ is deterministically zero and

\[
p(y \mid \gamma = 0)
= (2\pi\sigma^2)^{-n/2}
\exp\left(
-\frac{1}{2\sigma^2}
\sum_{i=1}^n y_i^2
\right).
\]

\noindent If $\gamma = 1$ then $\beta$ has prior distribution $\mathcal N(0,\sigma^2 g)$. Integrating $\beta$ out of the likelihood yields,

\[
p(y \mid \gamma = 1)
=
(2\pi\sigma^2)^{-n/2}
(1 + gn)^{-1/2}
\exp\left(
-\frac{1}{2\sigma^2}
\left[
\sum_{i=1}^n y_i^2
-
\frac{g}{1 + gn} S_1^2
\right]
\right).
\]

\noindent The posterior odds are given by

\[
\log\frac{\Pr(\gamma=1\mid y)}{\Pr(\gamma=0\mid y)}
=
\log\frac{\pi}{1-\pi}
-
\frac{1}{2}\log(1 + gn)
+
\frac{g}{2\sigma^2(1 + gn)} S_1^2.
\]
\noindent The posterior inclusion probability follows from applying the logistic transform to this log odds expression. Given $\gamma=1$, the posterior for the shared coefficient $\beta$ is Normal by conjugacy:

\[
\beta \mid y,\gamma=1,\sigma^2,g \sim \mathcal N(\mu^\star, v^\star),
\qquad
v^\star = \frac{\sigma^2}{n + 1/g},
\qquad
\mu^\star = \frac{n}{n + 1/g}\,\bar y.
\]

\noindent For scaling coefficients, which are always active, the same conjugacy with an uninformative prior gives

\[
\beta \mid y \sim \mathcal N(\bar y, \sigma^2/n).
\]

\noindent The inclusion probabilities $\pi$ are assigned a Beta prior, $\mathrm{Beta}(\tau_\pi m, \tau_\pi (1-m))$ where $m = \kappa_\pi 2^{-c_2 j}$. Conditioning on the inclusion indicators ${\gamma_i}$ for a wavelet level, the posterior for $\pi$ is given by
\[
\pi \mid \{\gamma_i\}
\sim
\mathrm{Beta}(\tau_\pi m + n_1,\; \tau_\pi(1 - m) + n_0),
\]
where $n_1 = \sum_i \gamma_i$ counts the number of included coefficients, $d$ is the total number of coefficients at this wavelet level, and $n_0 = d - n_1$ is the number excluded. This result follows from Beta-Binomial conjugacy.

Next, consider the slab variance parameter $g$. Assume that $n_{\mathrm{act}}$ coefficients are active at this level, with corresponding sampled values $\beta_1, \ldots, \beta_{n_{\mathrm{act}}}$. By Normal-Inverse-Gamma conjugacy,

\[
g \mid \{\beta_i\},\sigma^2
\sim
\mathrm{Inv\text{-}Gamma}
\left(
a_g + \frac{n_{\mathrm{act}}}{2},\;
b_g + \frac{1}{2}
\sum_{i=1}^{n_{\mathrm{act}}}
\frac{\beta_i^2}{\sigma^2}
\right).
\]

\noindent With the shared coefficient $\beta$, define the residuals as $r_i = y_i - \beta$. Let $N$ represent the total number of residual terms across all curves and coefficients for the given dimension. The prior for $\sigma^2$ is $\mathrm{Inv\text{-}Gamma}(a_\sigma, b_\sigma \tau)$. By the Normal--Inverse-Gamma conjugacy for a Normal likelihood with known mean, the posterior update is given by

\[
\sigma^2 \mid r,\tau
\sim
\mathrm{Inv\text{-}Gamma}
\left(
a_\sigma + \frac{N}{2},\;
b_\sigma \tau + \frac{1}{2}\sum_{i=1}^{N} r_i^2
\right).
\]

\noindent Finally, each dimension-specific variance $\sigma_m^2$ is assigned the prior $\mathrm{Inv\text{-}Gamma}(a_\sigma, b_\sigma \tau)$. Conditional on $\tau$, conjugacy in the Gamma--Inverse-Gamma hierarchical model yields:

\[
\tau \mid \{\sigma_m^2\}
\sim
\mathrm{Gamma}
\left(
a_\tau + D a_\sigma,\;
b_\tau + b_\sigma \sum_{m=1}^D \frac{1}{\sigma_m^2}
\right).
\]

\noindent These conditional distributions define the full Gibbs updates for the wavelet coefficients. Algorithm \ref{alg:wavelet-gibbs} presents the full Gibbs sampling procedure for posterior updating.

\subsection{Kernel Posterior Updates}
\label{sec:kernel_updates}

We now go through each kernel included in the model and write out its proposal density. The squared exponential kernel \citep{rasmussen2006gaussian}
\[
k_{\mathrm{SE}}(r;\ell)=\exp\!\left(-\frac{r^2}{2\ell^2}\right)
\]
contains a single lengthscale, which is proposed as
\[
\log\ell' \mid \log\ell \sim \mathcal{N}(\log\ell,\,\sigma_\ell^2),
\qquad
\ell' = \exp(\log\ell').
\]

The  Matérn–$3/2$ and Matérn–$5/2$ kernels \citep{rasmussen2006gaussian},
\[
k_{\mathrm{Mat}\,3/2}(r;\ell)
=
\big(1+\sqrt{3}r/\ell\big)\exp(-\sqrt{3}r/\ell),
\qquad
k_{\mathrm{Mat}\,5/2}(r;\ell)
=
\Big(1+\sqrt{5}r/\ell+\frac{5r^2}{3\ell^2}\Big)\exp(-\sqrt{5}r/\ell),
\]
also contain single lengthscales, and use the same proposal. 

To model periodic structure, we include the periodic kernel \citep{rasmussen2006gaussian}
\[
k_{\mathrm{Per}}(t,t';\ell,\rho)
=
\exp\!\left[-\frac{2\sin^2(\pi(t-t')/\rho)}{\ell^2}\right],
\]
with positive lengthscale $\ell$ and period $\rho\in(0,1)$ after rescaling time to $[0,1]$. The lengthscale is proposed on the log scale,
\[
\log\ell' \mid \log\ell \sim \mathcal{N}(\log\ell,\,\sigma_\ell^2),
\qquad
\ell' = \exp(\log\ell'),
\]
and the period is proposed on the logit scale,
\[
\operatorname{logit}(\rho') \mid \operatorname{logit}(\rho)
\sim \mathcal{N}(\operatorname{logit}(\rho),\,\sigma_\rho^2),
\qquad
\rho' = \operatorname{logit}^{-1}(\operatorname{logit}(\rho')).
\]

The rational quadratic kernel \citep{rasmussen2006gaussian}
\[
k_{\mathrm{RQ}}(r;\ell,\alpha)
=
\left(1 + \frac{r^2}{2\alpha\ell^2}\right)^{-\alpha}
\]
includes a lengthscale and a positive shape parameter $\alpha$, each updated as
\[
\log\ell' \mid \log\ell \sim \mathcal{N}(\log\ell,\,\sigma_\ell^2),\qquad
\ell' = \exp(\log\ell'),
\]
\[
\log\alpha' \mid \log\alpha \sim \mathcal{N}(\log\alpha,\,\sigma_\alpha^2),
\qquad
\alpha' = \exp(\log\alpha').
\]

The powered exponential kernel \citep{diggle1998model}
\[
k_{\mathrm{PowExp}}(r;\ell,\kappa)
=
\exp\!\left[-\left(\frac{r}{\ell}\right)^\kappa\right], \qquad \kappa\in(0,2],
\]
uses
\[
\log\ell' \mid \log\ell \sim \mathcal{N}(\log\ell,\,\sigma_\ell^2),
\qquad
\ell' = \exp(\log\ell'),
\]
and updates $\kappa$ through a logit transform,
\[
\operatorname{logit}\!\left(\frac{\kappa'}{2}\right)
\mid
\operatorname{logit}\!\left(\frac{\kappa}{2}\right)
\sim \mathcal{N}\!\left(\operatorname{logit}\!\left(\frac{\kappa}{2}\right),\,\sigma_\kappa^2\right),
\qquad
\kappa' = 2\,\operatorname{logit}^{-1}\!\left(\operatorname{logit}\!\left(\frac{\kappa'}{2}\right)\right).
\]

To capture nonstationary structure, we include a Gibbs kernel with piecewise–constant lengthscale in three equal time regions \citep{paciorek2003nonstationary}. Let $\ell(t)\in\{\ell_1,\ell_2,\ell_3\}$ denote the segment–specific lengthscale. The kernel
\[
k_{\mathrm{GibbsPC3}}(t,t')
=
\sqrt{\frac{2\,\ell(t)\,\ell(t')}{\ell(t)^2 + \ell(t')^2}}\,
\exp\!\left[
-\frac{(t-t')^2}{\ell(t)^2 + \ell(t')^2}
\right]
\]
is updated by proposing each $\ell_j$ as
\[
\log\ell_j' \mid \log\ell_j \sim \mathcal{N}(\log\ell_j,\,\sigma_{\ell_j}^2),
\qquad
\ell_j' = \exp(\log\ell_j'),
\qquad j\in\{1,2,3\}.
\]

Finally, we include a changepoint Matérn–$5/2$ kernel. As described by  \citet{lloyd2014automatic}, changepoint kernels are defined using additions and multiplications of sigmoidal functions. Thus we can define
\[
\sigma(t)=\big(1+\exp(-(t-t_0)/\delta)\big)^{-1},
\qquad
t_0\in(0,1),\ \delta>0,
\]
and let $\ell_{\mathrm{L}}$ and $\ell_{\mathrm{R}}$ denote the left and right lengthscales. The covariance is then
\[
k_{\mathrm{CP\text{-}M52}}(t,t')
=
(1-\sigma(t))(1-\sigma(t'))\,k_{\mathrm{Mat}\,5/2}(r;\ell_{\mathrm{L}})
+
\sigma(t)\sigma(t')\,k_{\mathrm{Mat}\,5/2}(r;\ell_{\mathrm{R}}).
\]
The parameters $\ell_{\mathrm{L}}$, $\ell_{\mathrm{R}}$, and $\delta$ are updated as
\[
\log\ell_{\mathrm{L}}' \mid \log\ell_{\mathrm{L}} \sim \mathcal{N}(\log\ell_{\mathrm{L}},\,\sigma_{\ell_{\mathrm{L}}}^2),\qquad
\ell_{\mathrm{L}}'=\exp(\log\ell_{\mathrm{L}}'),
\]
\[
\log\ell_{\mathrm{R}}' \mid \log\ell_{\mathrm{R}} \sim \mathcal{N}(\log\ell_{\mathrm{R}},\,\sigma_{\ell_{\mathrm{R}}}^2),\qquad
\ell_{\mathrm{R}}'=\exp(\log\ell_{\mathrm{R}}'),
\]
\[
\log\delta' \mid \log\delta \sim \mathcal{N}(\log\delta,\,\sigma_\delta^2),
\qquad
\delta' = \exp(\log\delta'),
\]
and the changepoint location is updated via a logit proposal
\[
\operatorname{logit}(t_0') \mid \operatorname{logit}(t_0)
\sim \mathcal{N}(\operatorname{logit}(t_0),\,\sigma_{t_0}^2),
\qquad
t_0' = \operatorname{logit}^{-1}(\operatorname{logit}(t_0')).
\]

\subsection{ICM Posterior Updates}
\label{app:icmupdates}

After the kernel family and its hyperparameters are updated, the parameters $(\tau_{B,k},B_k^{\text{shape}},\eta_k)$ are sampled using Metropolis–Hastings steps. The amplitude parameter $\tau_{B,k}$ is updated on the log scale using a Gaussian random walk proposal
\[
\log\tau_{B,k}' = \log\tau_{B,k} + \epsilon_\tau,\qquad \epsilon_\tau \sim \mathcal{N}(0,s_\tau^2).
\]
Each dimension-specific nugget $\eta_{k,m}$ is updated independently in the same manner,
\[
\log\eta_{k,m}' = \log\eta_{k,m} + \epsilon_{\eta,m},\qquad \epsilon_{\eta,m} \sim \mathcal{N}(0,s_\eta^2).
\]
The shape matrix $B_k^{\text{shape}}$ is updated by perturbing its Cholesky factor. Writing $B_k^{\text{shape}} = D\,LL^\top/\mathrm{tr}(LL^\top)$, we propose
\[
L' = L + E,
\]
where $E$ is lower-triangular with independent Gaussian entries $E_{ab}\sim\mathcal{N}(0,s_B^2)$ for $a\ge b$.

\section{Convergence Diagnostics}
\label{app:con}

In this appendix, we provide representative convergence diagnostics for the proposed MCMC algorithm. Due to the high dimensionality of the parameter space, it is not feasible to assess mixing for all parameters individually. Instead, we focus on a subset of informative quantities, including the Dirichlet process concentration parameter $\alpha$, the log-likelihood, and clustering-related summaries.

\begin{figure}[htbp]
    \centering
    
    \begin{subfigure}{0.48\textwidth}
        \centering
        \includegraphics[width=\linewidth]{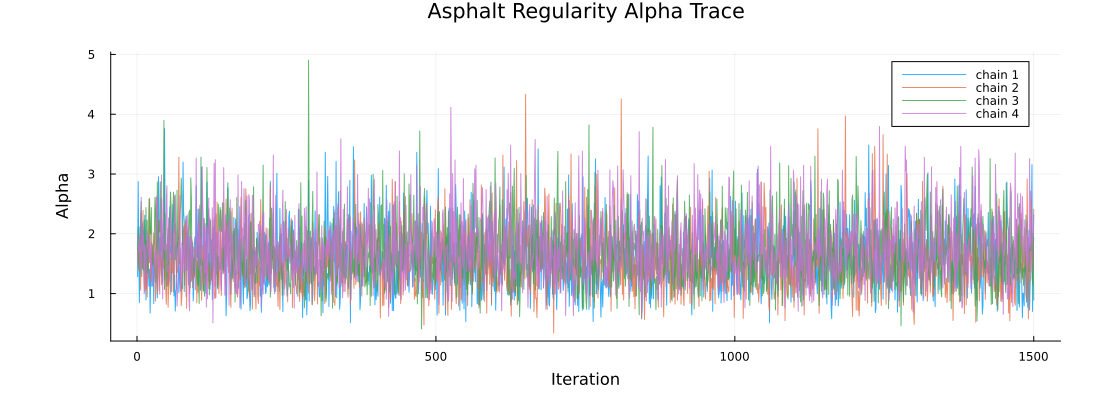}
        \caption{Alpha trace}
    \end{subfigure}
    \hfill
    \begin{subfigure}{0.48\textwidth}
        \centering
        \includegraphics[width=\linewidth]{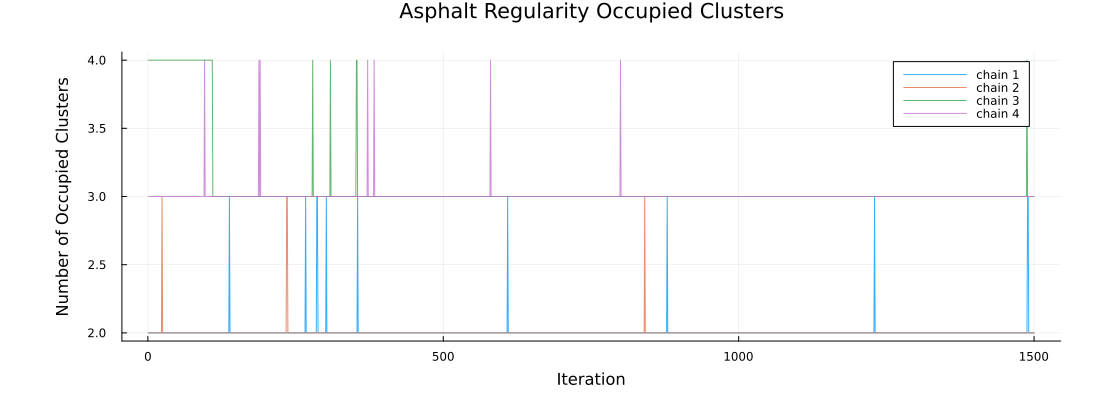}
        \caption{Number of clusters}
    \end{subfigure}
    
    \vspace{0.5cm}
    
    \begin{subfigure}{0.48\textwidth}
        \centering
        \includegraphics[width=\linewidth]{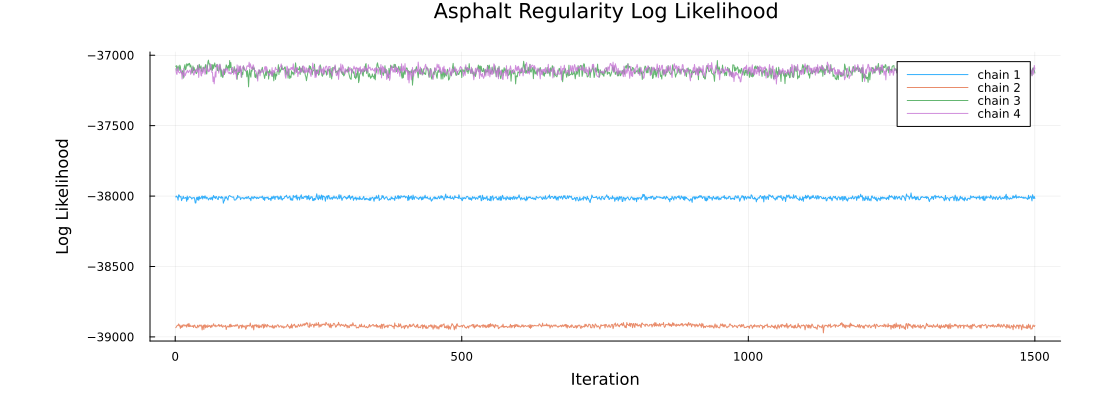}
        \caption{Log-likelihood}
    \end{subfigure}
    \hfill
    \begin{subfigure}{0.48\textwidth}
        \centering
        \includegraphics[width=\linewidth]{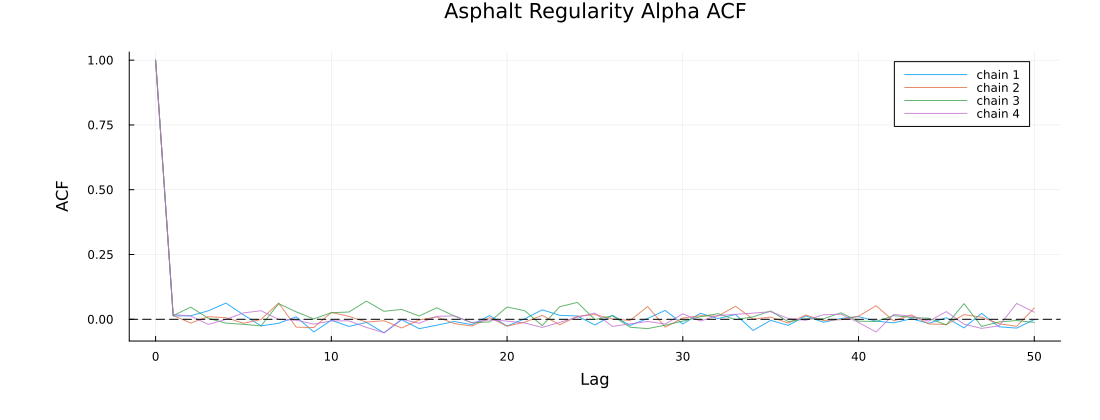}
        \caption{Alpha ACF}
    \end{subfigure}
    
    \vspace{0.5cm}
    
    \begin{subfigure}{0.48\textwidth}
        \centering
        \includegraphics[width=\linewidth]{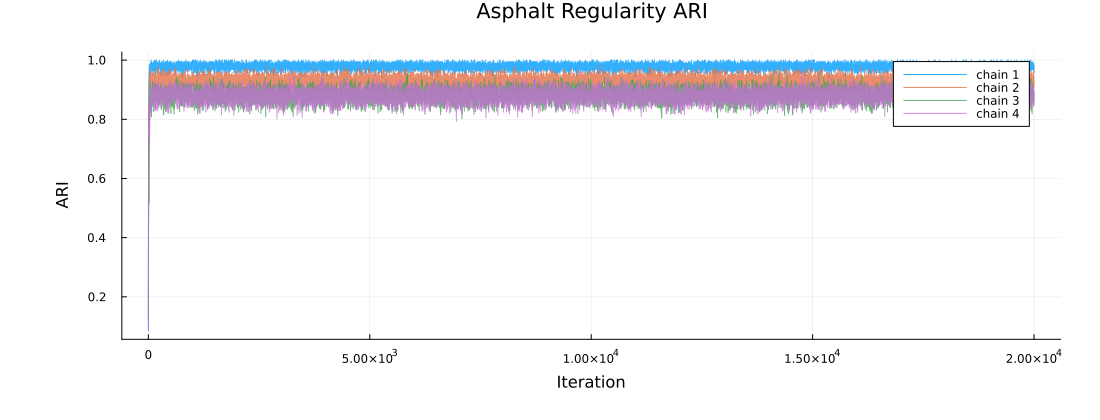}
        \caption{ARI vs previous iteration}
    \end{subfigure}
    \hfill
    \begin{subfigure}{0.48\textwidth}
        \centering
        \includegraphics[width=\linewidth]{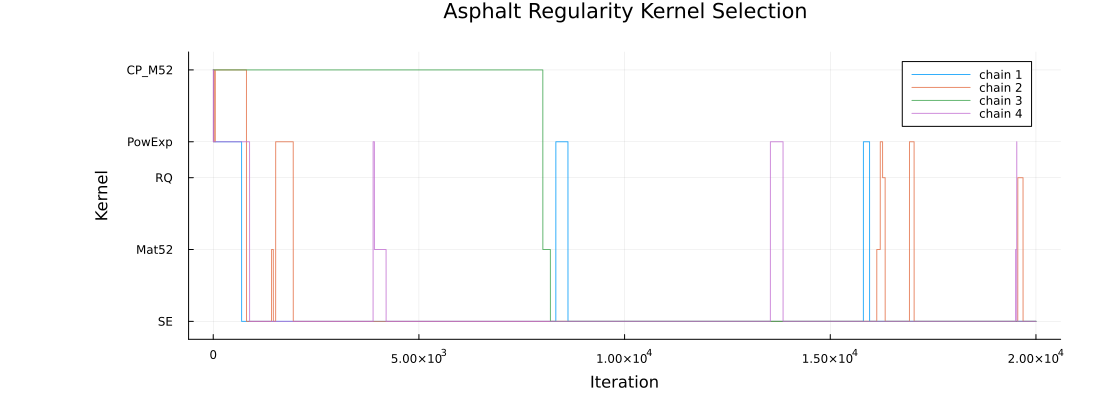}
        \caption{Dominant kernel}
    \end{subfigure}
    
    \caption{
MCMC convergence diagnostics for the Asphalt Regularity dataset across multiple chains.
(a) Trace plot of the Dirichlet process concentration parameter $\alpha$.
(b) Number of occupied clusters over iterations.
(c) Log-likelihood trajectories, illustrating variability across chains due to posterior multimodality.
(d) Autocorrelation function (ACF) of $\alpha$, showing rapid decay.
(e) Adjusted Rand index (ARI) between successive iterations, indicating clustering stability after burn-in.
(f) Dominant kernel selection over time, with the squared exponential kernel most frequently selected.
}
    \label{fig:six_subfigures}
\end{figure}

Figure~\ref{fig:six_subfigures} summarizes these diagnostics across multiple chains. Panels (a) and (d) display the trace and autocorrelation function (ACF) for $\alpha$, respectively. These plots indicate good mixing behavior, with rapid decay in autocorrelation and strong agreement across chains, suggesting that the sampler adequately explores this component of the posterior.

In contrast, panel (c) shows that the log-likelihood trajectories differ across chains, reflecting the highly multimodal nature of the posterior distribution. This behavior highlights the importance of running multiple chains and selecting solutions corresponding to high-likelihood regions, as the sampler may become trapped in local modes. This multimodality is further reflected in panel (b), where the number of occupied clusters varies substantially across iterations and chains.

Panel (f) shows the dominant kernel selected within each cluster over time. Across chains, the squared exponential kernel is most frequently selected, suggesting that smooth covariance structure provides the best fit for this dataset. Finally, panel (e) displays the adjusted Rand index (ARI) between successive clustering assignments. After the burn-in period, ARI values remain consistently high, indicating that cluster assignments stabilize and the inferred partition is relatively robust.

\newpage
\bibliography{sample}

\end{document}